# Cation accumulation drives the preferential partitioning of DNA in an aqueous two-phase system

Hiroki Sakuta[1,2†], Yuki Akamine[1†], Akari Kamo[3], Hao Gong[1], Norikazu Ichihashi[1,2], Arash Nikoubashman[4,5,6*], Miho Yanagisawa[1,2,3*]

1 Komaba Institute for Science, Graduate School of Arts and Sciences, The University of Tokyo, Komaba 3-8-1, Meguro, Tokyo 153-8902, Japan
2 Center for Complex Systems Biology, Universal Biology Institute, The University of Tokyo, Komaba 3-8-1, Meguro, Tokyo 153-8902, Japan
3 Department of Physics, Graduate School of Science, The University of Tokyo, Hongo 7-3-1, Bunkyo, Tokyo 113-0033, Japan
4 Leibniz-Institut für Polymerforschung Dresden e.V., Hohe Straße 6, 01069 Dresden, Germany
5 Institut für Theoretische Physik, Technische Universität Dresden, 01069 Dresden, Germany
6 Department of Mechanical Engineering, Keio University, Hiyoshi 3-14-1, Kohoku, Yokohama 223-8522, Japan

**AUTHOR INFORMATION**

†H.S. and Y.A. contributed equally to this paper.

**Corresponding authors**

E-mail: anikouba@ipfdd.de; myanagisawa@g.ecc.u-tokyo.ac.jp

**Abstract**

Mixtures of polyethylene glycol (PEG) and dextran (Dex) represent a widely used class of aqueous two-phase systems (ATPS), with applications ranging from the purification of various biomolecules such as nucleic acids to the synthesis of protocells. A key feature underlying these applications is the selective accumulation of biomolecules within Dex-rich droplets in an aqueous PEG phase, but the physical origin of this partitioning remains unclear. Entropic interactions were long assumed to be the primary driving force; however, our systematic experiments using DNA of different lengths indicate that entropy alone cannot fully explain the observed behavior. We identify an additional and previously underappreciated contribution from electrostatic interactions: Dex carries a slightly more negative charge than PEG, which drives preferential cation accumulation in the Dex-rich phase. These counterions facilitate the selective partitioning of DNA inside the Dex-rich droplets. This mechanism explains the dependency of DNA uptake in Dex-rich droplets on polymer length and salt concentration. Our findings establish that Donnan-type ion partitioning plays a crucial role in the localization of long nucleic acids in Dex-rich droplets, offering a unified explanation for this long-standing phenomenon. They lay the foundation for designing ATPS-based systems and help elucidate the physicochemical principles of biomolecular partition upon phase separation in cells.

**Main Text**

Aqueous two-phase systems (ATPS), such as mixtures of polyethylene glycol (PEG) and dextran (Dex), are classical examples of polymer systems exhibiting liquid–liquid phase separation (LLPS)[1-2]. Owing to their high biocompatibility and ease of preparation using commercially available polymers of varying molecular weights, PEG/Dex mixtures have found extensive applications in biotechnology, including cell sorting and biomolecule purification[3-6].

The membrane-free surface of the phase-separated, μm-sized Dex-rich droplets allows selective exchange and concentration of functional molecules, including proteins, enzymes, RNA, and notably DNA,[7-9] which can enhance diverse biochemical reactions[10-14], such as nucleic acid reactions, and transcriptional and regulatory complex assembly. These features make ATPS promising platforms for exploring biomedical applications[15], intracellular organization[16-17], and designing synthetic protocells[17-19].

Despite these recent advances, a long-standing question concerns the physical principles that govern how biomolecules selectively partition into the Dex-rich phase. In particular, previous experiments observed that long DNA strands accumulate in the Dex-rich phase[7, 9, 20], which appears counter-intuitive at first glance, since such a crowded and confined environment should impose significant conformational penalties[21]. Although entropy and electrostatic interactions have been proposed as mechanisms driving this behavior[6, 22-23], a comprehensive and systematic framework that fully accounts for the observed DNA partitioning is still missing.

To understand this peculiar behavior, we systematically investigated the localization of double-stranded DNA (dsDNA) of different lengths in PEG/Dex mixtures. Estimates based on the Asakura-Oosawa model suggest that the depletion pressure from small PEG chains in the aqueous phase pushes long dsDNA strands into the Dex-rich droplets. However, such calculations must be treated with caution for open and flexible objects such as polymers; indeed, control experiments with smaller depletants led to weaker dsDNA accumulation, contradicting the theoretical expectation. This discrepancy highlights the need for considering additional interactions. Therefore, we analyzed the cation partition ratio, finding a notable excess of cations in the Dex-rich phase, likely originating from the slightly negative charge of Dex in salt-containing buffers. This cation surplus then promotes the aggregation of dsDNA into the Dex-rich phase, due to the negatively charged sugar-phosphate backbone of dsDNA. These findings provide new insights into broader principles of molecular organization in phase-separated environments. They lay the foundation for designing ATPS-based systems in synthetic biology and soft matter and help elucidate the physicochemical principles of LLPS in cells.

**Results**

To analyze the partition ratio of dsDNAs between Dex-rich droplets and the aqueous PEG phase, we mixed 10 ng/μL dsDNA stained with SYBR Green I with PEG/Dex solutions (Fig. 1a). The PEG/Dex solutions consist of 6 wt% PEG6k and 2 wt% Dex500k, which lies within the miscibility gap, resulting in the formation of Dex-rich droplets[24-25]. The buffer used here is 8 mM Tris-HCl with 4% 2-Mercaptoethanol (see Experimental Section and S1, 2 in Supplemental Information (SI) for the details). The presence of salts in the PEG/Dex mixtures makes it challenging to infer the exact concentrations of each phase from density and refractive index measurements. Roughly estimated compositions are 12 wt% Dex and 1 wt% PEG for the Dex droplets and 6 wt% PEG and 2 wt% Dex for the surrounding PEG-rich phase (see S3 and Fig. S1).

First, we use long ~49 kbp λ DNA with an estimated radius of gyration $R_g$ ~500 nm[26-27], which is much larger than its persistence length (~50 nm[26]) and the average mesh size of chains in the Dex-rich droplets (~12 nm) (see S4 for the estimation). Transmitted light images alongside confocal microscopy images of a cross-section of Dex-rich droplets in an aqueous PEG phase (red) clearly demonstrate an accumulation of dsDNA (green) within the Dex-rich droplets (Figs. 1b, 1c). The 3D cross-sectional image and the intensity profile along the *z*-axis further corroborate that the dsDNA strands are distributed within the Dex droplets (Fig. 1d), as reported previously.[9] To investigate how the length of dsDNA affects its accumulation within Dex-rich droplets, we repeated similar experiments using eight different strand lengths, approximately 49 kbp (λ DNA), 24 kbp, 10 kbp, 5 kbp, 2 kbp, 0.2 kbp, 0.1 kbp, and 0.05 kbp (see also Fig. S2). Except for the shortest strands with 0.05 kbp, we observed a notable increase in dsDNA accumulation in the Dex-rich droplets (Fig. 1e and Fig. S3).

To quantify the length-dependent accumulation of dsDNA, we analyze the fluorescence intensity difference between the Dex-rich and aqueous PEG phases, $I_{Dex} - I_{PEG}$. To eliminate curvature effects, the average intensity of dsDNA was measured within the droplets, excluding ~1 μm from the surface (Fig. 1f, see Experimental Section). The intensity difference was normalized by a constant $I_0$, measured from polymer-free dsDNA solutions, $\Delta I$ (= $(I_{Dex} - I_{PEG})/ I_0$), and plotted against the dsDNA length. Figure 1g shows that $\Delta I$ initially increases as the dsDNA length increases from 0.05 kbp to 2 kbp and then reaches a plateau for longer strands. This length-dependent accumulation of dsDNA was also confirmed in experiments performed at constant molar concentration (Fig. S4). Notably, the inflection point of $\Delta I$ is located near 0.1-0.15 kbp, which corresponds to the persistence length of dsDNA (~50 nm)[26]. Thus, the most significant increase in accumulation occurs for dsDNA that resembles almost rigid rods. In this regime, each additional base pair increases the net charge without significantly changing the conformational entropy, which could explain the steep increase of $\Delta I$ for short dsDNA.

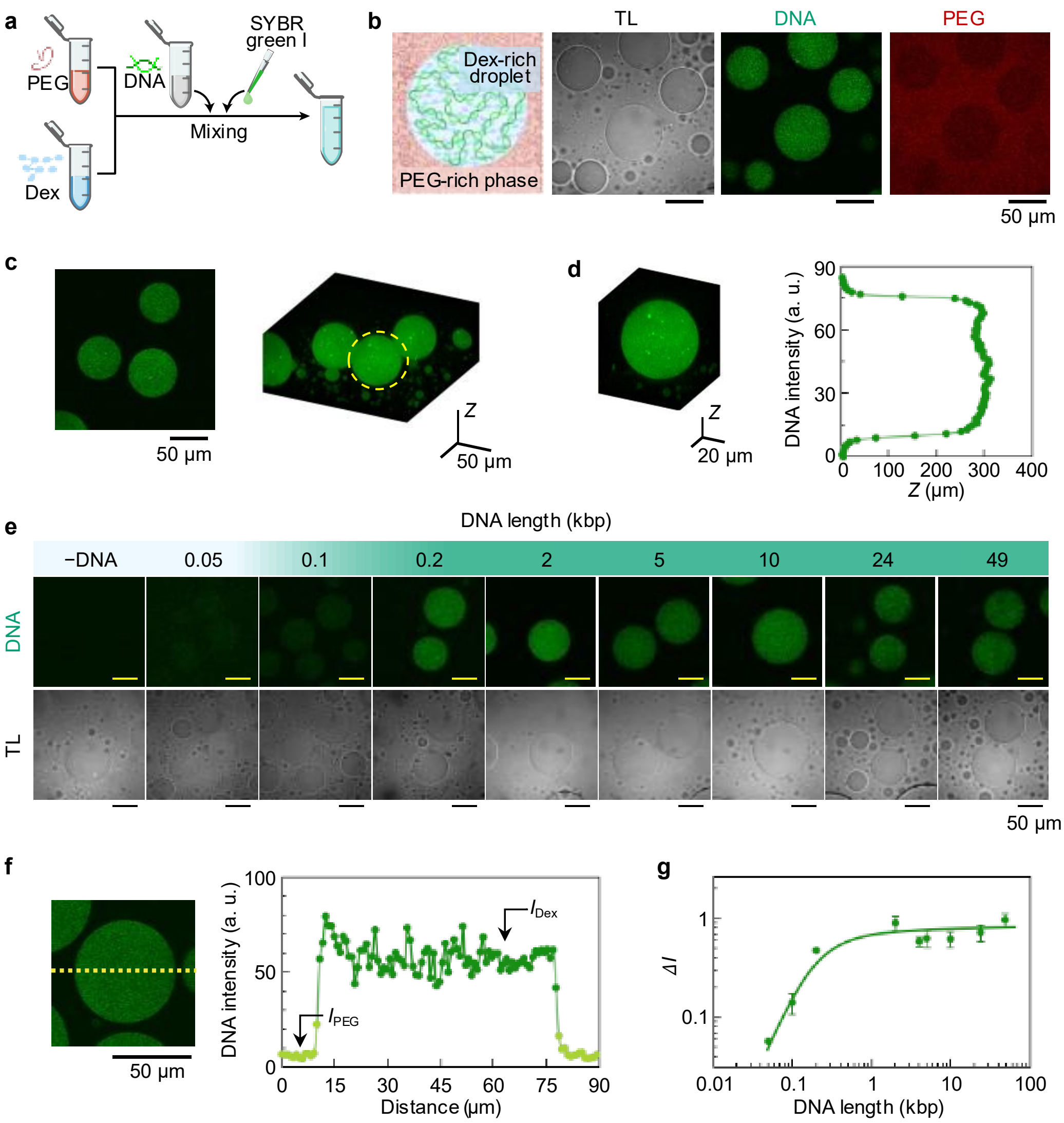

**Figure 1. Length-dependent accumulation of dsDNA within Dex-rich droplets in an aqueous PEG phase.** (a) Schematic representation of the preparation process illustrating Dex500k droplets (blue) containing dsDNA stained with SYBR Green I (green), suspended in a PEG6k phase with 0.2 wt% RB-PEG5k (red). (b) Transmitted light (TL) image alongside confocal microscopy images of a cross-section of Dex droplets with ~49 kbp λ DNA (green) in the PEG phase (red). (c) Cross-sectional fluorescence image of dsDNA and the 3D view. (d) The 3D image displays the Dex-rich droplet, outlined by a dotted line in (c), along with its fluorescence intensity profile along the *z*-axis. (e) Confocal fluorescence microscopy (top) and TL images (bottom) for varying dsDNA lengths in Dex-rich droplets. (f) Fluorescence intensity profile along the dotted line on the left. (g) Intensity difference between Dex-rich and aqueous PEG phases, normalized by a constant $I_0$ derived from the value of λ DNA without polymers: $\Delta I = (I_{\mathrm{Dex}} - I_{\mathrm{PEG}}) / I_0$. The error bars represent the standard error ($n \geq 35$ for each length). The solid green line is a visual guide. The $\Delta I$ without dsDNA is approximately null.

Two mechanisms have been brought forward to explain why dsDNA strongly partitions into the Dex-rich phase, namely entropic contributions (including both depletion forces and polymer mixing entropy), and electrostatic interactions between these components[6, 22-23]. We first estimated the contribution from depletion interactions. Treating the dsDNA strands as worm-like chains with a Kuhn length of 100 nm, we estimate $R_g$~500 nm for ~49 kbp λ DNA, which is much larger than $R_{g,PEG}$~3 nm and $R_{g,Dex}$~24 nm estimated for PEG and Dex, respectively[28] (see also the experimentally obtained hydrodynamic diameters, Fig. S5). Within the Asakura-Oosawa (AO) model framework (see S4 for further details), this size asymmetry leads to a depletion-induced attraction between dsDNA coils of $\Delta U_{PEG} \sim -29 k_B T$ in the aqueous PEG phase, which is considerably stronger than $\Delta U_{Dex} \sim -10 k_B T$ calculated for the Dex-rich droplet. Thus, placing dsDNA chains inside the Dex-rich droplets incurs a *smaller* entropic penalty to the overall free energy, which could explain the experimentally observed partitioning.

To test the importance of depletion interactions, we tried to enhance the underlying entropic driving force by replacing 6 wt% PEG6k with 16 wt% PEG1k in the PEG/Dex mixtures. Although this replacement increases the critical concentration from ~8 wt% to ~16 wt%, the estimated Dex concentration in Dex-rich droplets is almost similar (~20 wt%, see S3). Strikingly, this replacement significantly *weakened* dsDNA localization (Fig. 2a), despite the AO model predicting that smaller depletants should enhance depletion forces. This discrepancy points to the presence of additional interactions and also highlights limitations of AO-type estimates, which assume that PEG and Dex act as mutually impenetrable hard spheres relative to dsDNA. In reality, dsDNA is a highly open and sparsely filled coil, which – in the case of ~49 kbp λ DNA – occupies only about 0.006% of the spherical volume defined by its $R_g$.

In addition to depletion, the entropy of polymer mixing could also explain dsDNA partitioning. Flory-Huggins theory predicts that it is entropically favorable to insert the dsDNA chains into the Dex-rich droplet phase rather than into the surrounding aqueous phase, since replacing the larger Dex molecules removes fewer translational degrees of freedom from the system (see S5). However, within this mean-field description, the entropic contribution to the insertion free energy depends only weakly on chain length, which is at odds with the steep increase of DNA accumulation observed in our experiments (Fig. 1g). By contrast, an enthalpic contribution (*i.e.*, a non-zero $\chi$ parameter for the dsDNA-droplet interactions) could explain the almost linear initial increase in accumulation with dsDNA length.

To assess the influence of electrostatic interactions, we investigated the phase partitioning of ~49 kbp λ DNA in pure water and in NaCl solutions at various concentrations, instead of an 8 mM Tris-HCl buffer. As shown in Fig. 2b, nearly all dsDNA strands are located within Dex-rich droplets at NaCl concentrations of 0 mM and 100 mM. However, at a NaCl concentration of 1 M, this localization is significantly reduced, which is likely due to the strong screening of electrostatic interactions between dsDNA and Dex or PEG. Note, however, that some counterions are attached to the dsDNA prior to solvation, so the *overall* salt concentration in the 0 mM NaCl solution is *not* perfectly zero. Calibration of the $Na^+$ probe with and without dsDNA (Fig. S6), together with the estimated $Na^+$ concentration contributed by dsDNA-associated counterions (~30 μM), indicates that the intrinsic counterions of dsDNA are negligible and do not affect the NaCl-dependent dsDNA localization. When NaCl was replaced with $MgCl_2$, dsDNA localization within Dex-rich droplets was suppressed already at ~20 μM $MgCl_2$, much lower than the concentration required for NaCl (Figs. 2c, 2d). This behavior suggests that multivalent cations strongly enhance the electrostatic interactions between dsDNA and Dex or PEG.

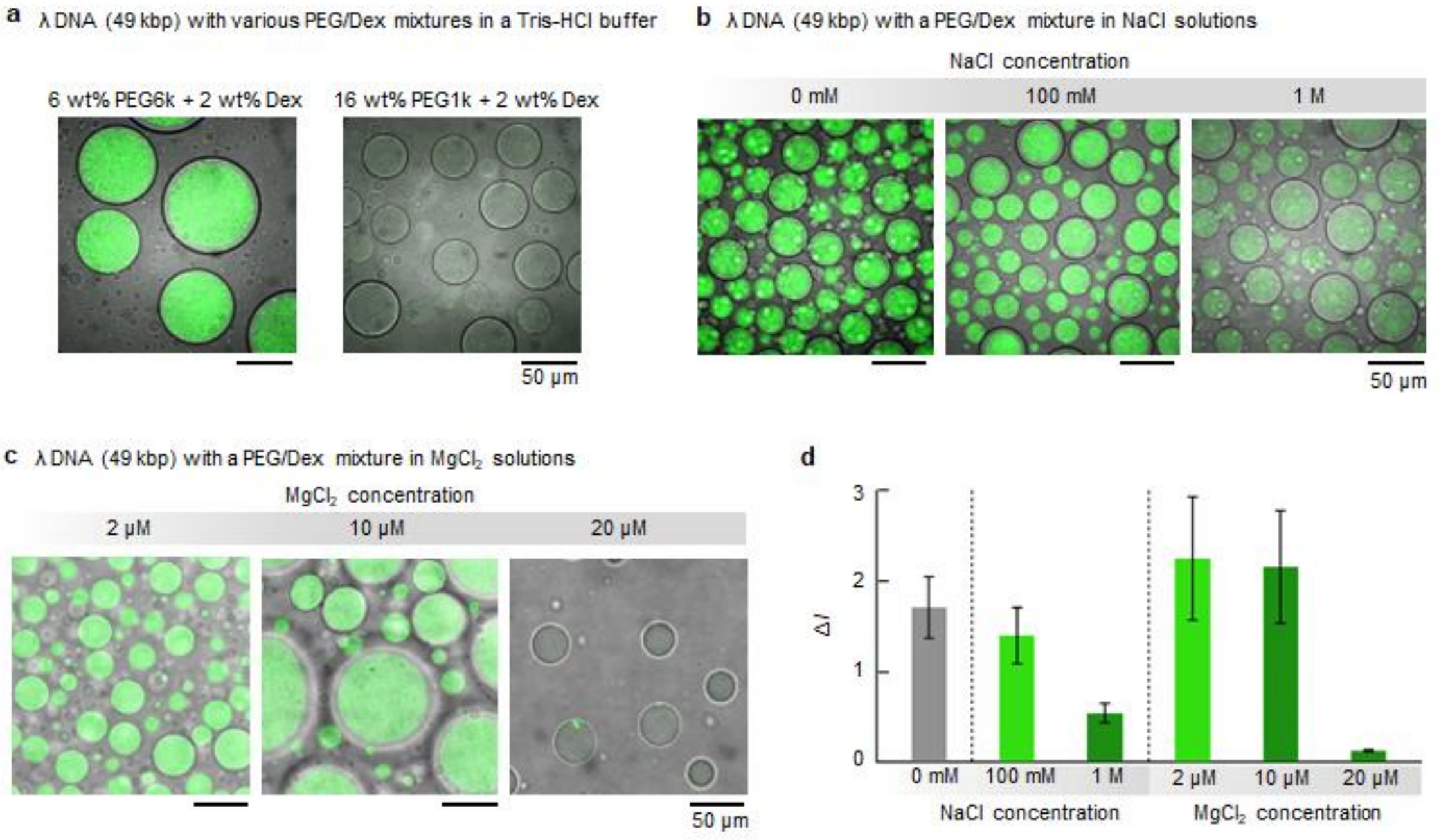

**Figure 2. Changes in the degree of dsDNA accumulation inside Dex-rich droplets by modifying the solution conditions.** Merged images show the TL image of Dex-rich droplets in an aqueous PEG phase along with the fluorescent image of ~49 kbp λ DNA (green). (a) dsDNA was added to the mixtures consisting of 2 wt% Dex500k and either 6 wt% PEG6k (left) or 16 wt% PEG1k (right) in 8 mM Tris-HCl buffer. (b, c) Instead of Tris-HCl buffers, 0 mM (purified water), 100 mM, and 1 M NaCl solution (b; left to right) or 2 μM, 10 μM, and 20 μM, $MgCl_2$ solution were used (left to right). It is essential to note that some counterions are bound to the dsDNA prior to solvation, which means that the overall salt concentration in the solutions is *not* precisely zero (see Fig. S7). (d) The normalized intensity difference $\Delta I$ for various solution conditions, (from left to right) 0 mM, 100 mM, 1M NaCl, and 2 μM, 10 μM, and 20 μM $MgCl_2$. The error bars indicate standard deviation ($N = 3$).

To understand this behavior, we measured the ζ potential of the Dex and PEG chains under different solution conditions (see experimental section and S6 for the method). In pure water (see control data in Fig. 3a), Dex exhibited a more negative charge compared to PEG. The addition of salts like NaCl or $MgCl_2$ caused the ζ potentials of both Dex and PEG to become nearly zero, with Dex showing a more drastic change than PEG. This difference indicates that Dex, being more negatively charged in pure water, has a higher tendency to attract and bind cations than PEG.

Although oxygen atoms in PEG molecules may cause partial negative charges, their $pK_a$ values are typically above 14, indicating that PEG is essentially neutral in water. In contrast, Dex, a polysaccharide, readily forms hydrogen bonds via its numerous hydroxyl (-OH) and glycosidic groups. Its $pK_a$ is around 12,[29] making it moderately more acidic than PEG. This difference in acidity aligns with the observation that Dex is more negatively charged than PEG in water (Fig. 3a). As the molecular weight decreases, the ζ potential of Dex tends toward zero (Fig. S7), while that of PEG becomes slightly positive, likely due to the reduced number of oxygen atoms (Fig. 3a). Adding salt shifts the ζ potential of PEG from slightly positive (especially for PEG1k rather than PEG6k) to slightly negative, indicating some anion adsorption onto the PEG chain. This trend explains why dsDNA accumulation drastically decreases when PEG6k is replaced by PEG1k (Fig. 2a). In this case, the electrostatic interaction between dsDNA and the slightly positively charged PEG1k-rich phase may resemble its interaction with the cation enriched Dex500k phase. As a result, dsDNA strands are nearly equally distributed in both the PEG1k-rich and Dex500k-rich phases. In principle, direct or water-mediated interactions between dsDNA, PEG, and Dex (e.g., hydrogen bonding, differences in hydration) may also contribute to the observed partitioning. Given the highly hydrophilic nature of all three components, we surmise that such effects are unlikely to be dominant, but we cannot fully separate non-electrostatic contributions from electrostatics within the present experimental constraints.

To further characterize electrostatic interactions, we quantified the accumulation of cations using $Na^+$ probe in the PEG/Dex mixtures with NaCl (see experimental section and S7 for the details). When the $Na^+$ probe was added to PEG/Dex mixtures, it primarily localized in the PEG-rich aqueous phase in the absence of NaCl, suggesting a higher affinity of the $Na^+$ for PEG than for Dex. To eliminate the interaction between the polymers and the $Na^+$ probe and quantify the distribution of $Na^+$ in each phase, we decided to add the $Na^+$ probe after isolating the PEG and Dex phases from the phase-separated PEG/Dex solution after centrifugation (Fig. 3b).

In a control experiment, 1 M NaCl and $Na^+$ probe were added to both the PEG- and Dex-rich phases after they were isolated. The fluorescence intensity of the $Na^+$ probe from each solution was nearly identical (Fig. S8), suggesting that they contained the same salt concentration, as expected. However, when 100 mM NaCl was added *before* isolating the PEG- and Dex-rich phases, with or without λ DNA, the fluorescence intensity of the Dex-rich phase was significantly higher than that of the PEG-rich phase, the presence of DNA elevated the contrast (Fig. 3c). This trend remained consistent across various NaCl concentrations up to 1 M (Fig. 3d and Fig. S9), indicating that the Dex-rich phase contained more $Na^+$ compared to the PEG-rich phase. In experiments using the $Cl^-$ probe (see S7 for the details), the fluorescence intensity difference between the PEG-rich and Dex-rich phases was already present at 0 mM added NaCl and did not increase in the presence of DNA (Fig. 3e). Unlike the $Na^+$ probe, this difference originates from probe–polymer interactions, suggesting that $Cl^-$ ions distribute nearly evenly between the two phases. These results clearly demonstrate that adding NaCl to the PEG/Dex mixtures causes the cationic $Na^+$ to preferentially accumulate in the Dex-rich phase, which is more negatively charged in pure water (Fig. 3a).

Our results demonstrate that cation accumulation (or Donnan partitioning of counterions) plays a central role in driving dsDNA accumulation inside Dex-rich droplets. Additionally, the observed length dependence of dsDNA accumulation (Fig. 1g) aligns closely with the accumulation behaviors seen in charged polymer systems, where electrostatic interactions mainly drive LLPS.[30] These observations highlight the importance of including electrostatic interactions alongside ion partitioning in our understanding of LLPS mechanisms within charged polymer systems. ATPS droplets, including Dex-rich droplets, have been widely used as model systems for synthetic protocells and intracellular LLPS condensates. Since ion concentrations like $Na^+$ and $K^+$ are crucial for various biochemical reactions, our findings on the uneven distribution of cations in coexisting phases could greatly aid in designing and understanding ATPS and LLPS systems with appropriate compositions.

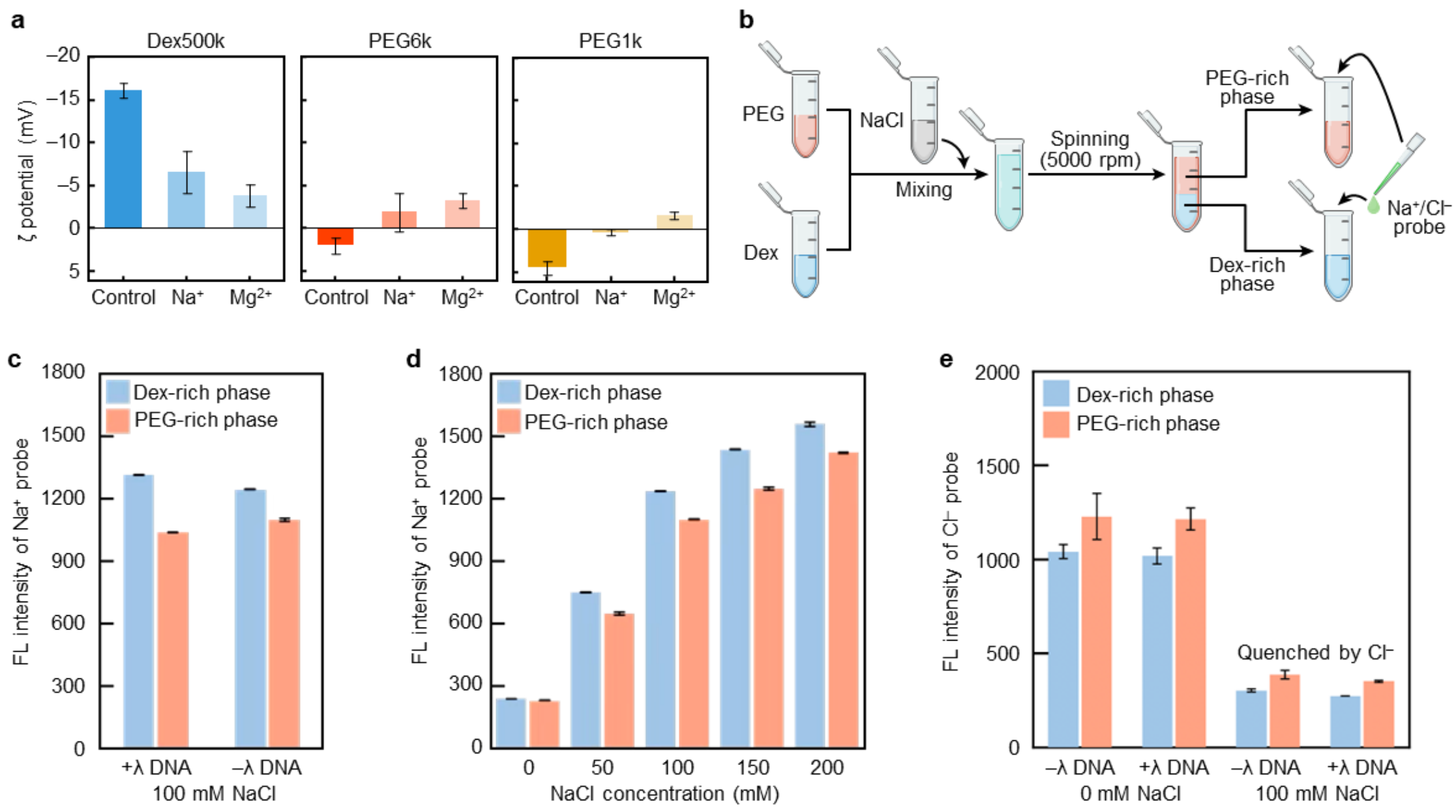

**Figure 3: Surface charge of PEG and Dex chains and enhanced cation accumulation in the Dex phase.** (a) ζ potential of 0.1wt% Dex500k, 0.1 wt% PEG6k, 0.5 wt% PEG1k in water (control), 200 mM NaCl ($Na^+$), and 200 mM $MgCl_2$ ($Mg^{2+}$). (b) $Na^+$ probe was added to the PEG and Dex phases with NaCl after isolating the two phase-separated liquids. (c, d) The $Na^+$ probe intensity (a. u.) in each PEG- and Dex-rich phase at various NaCl concentrations added before isolation. The solution in (c) contains 100 mM NaCl, with (+) or without (–) λ DNA. (e) The $Cl^-$ probe intensity (a. u.) in each PEG- and Dex-rich phase at 0 mM (purified water), 100 mM NaCl concentrations added before isolation the mixture with (+) or without (–) λ DNA. The error bars indicate standard deviation ($N = 3$ for (a, e) and 5 (c, d)). The control data, with 1 M NaCl added before and after the isolation, are shown in Figs. S8, S9.

## Experimental Section

To prepare binary polymer blends of polyethylene glycol (PEG) and dextran (Dex), we used Dex from *Leuconostoc* spp. with a molar weight (Mw) of 450–650 kg/mol (Dex500k; 31392; Sigma-Aldrich, St. Louis, MO, USA), polyethylene glycol with a Mw of 7–9 kg/mol (PEG6k; PEG6000; 169–09125; FUJIFILM Wako Pure Chemical Co., Osaka, Japan) or 900–1000 g/mol (PEG1k; 165-09085; FUJIFILM Wako Pure Chemical Co.), which dissolved in ultrapure distilled water (Invitrogen, Waltham, MA, USA). To visualize the PEG-rich phase with PEG6k, we added rhodamine-B-labeled methyl polyethylene glycol (5 kg/mol, RB-PEG; FL001045-5k, Biopharma

PEG, Watertown, MA, USA) with 0.2 wt%. NaCl (191-01665; FUJIFILM Wako Pure Chemical Co.) was used with the polymers to assess the influence of electrostatic interactions. Dex droplets were generated using mixtures of 6 wt% PEG6k and 2 wt% Dex500k, or 16 wt% PEG1k and 2 wt% Dex500k, yielding droplets with radii of 4–50 μm. We have confirmed that 100 mM NaCl barely altered the coexisting PEG-rich and Dex-rich phases (S3 and Fig. S1 in the SI).

Experiments were performed on DNA of lengths ranging from 50 bp to 49 kbp. λ DNA, as 49kbp DNA, was purchased from Nippon Gene Co., Ltd. (318-00414; Tokyo, Japan). As 50 bp, we used double-stranded DNA (dsDNA) fragments (NoLimits 50 bp DNA Fragment; SM1421; Thermo Scientific, Waltham, MA, USA). To obtain a 24 kbp fragment, λ DNA (New England Biolabs) was digested with XbaI (Takara Bio Inc.) at 37 °C for two hours. XbaI cleaves λ DNA at nucleotide position 24508, yielding fragments of 24508 bp and 23994 bp. 5 kbp, 2 kbp, 0.2 kbp, and 0.1 kbp DNA were synthesized by PCR using KOD One PCR Master Mix (KMM-101; Tokyobo Co., Ltd., Osaka, Japan), λ DNA (Klenow Fragment-treated; 318-05291; Nippon Gene Co., Ltd.) as a template, and primers (see S1 for the details). SYBR Green I Nucleic Acid Gel Stain (5761A; Takara Bio Inc., Shiga, Japan), 2-Mercaptoethanol (131-14572; FUJIFILM Wako Pure Chemical Co.), and Tris-HCl (pH 8.0; Nacalai Tesque, Kyoto, Japan) were used together with a binary polymer blend for DNA observation. The final concentrations were 6 wt% PEG (or 16 wt% PEG1k), 2 wt% Dex500k, 0.6X SYBR Green I, 8 mM Tris-HCl, 4% 2-Mercaptoethanol, and 10 ng/μL DNA.

Fluorescence images of the Dex droplets in a PEG solution were obtained using a confocal laser scanning microscope (IX83, FV1200; Olympus Inc., Tokyo, Japan), equipped with a water-immersion objective lens (UPLSAPO 60XW, Olympus Inc.). DNA localization was quantified by comparing the difference in fluorescence intensity between the inside and outside of the Dex-rich droplets. The average intensity of the PEG-rich and Dex-rich phases (droplets) and the intensity of λ DNA in buffer solution (bulk) were defined as $I_{\mathrm{PEG}}$, $I_{\mathrm{Dex}}$ and $I_0$, respectively. Then, the normalized intensity difference, $\Delta I = (I_{\mathrm{Dex}} - I_{\mathrm{PEG}})/I_0$, was analyzed.

ζ potential measurements were performed using a zeta potential meter with a flat electrode cell chamber (ELSZneo; Otsuka Electronics Co., Ltd., Osaka, Japan). Measurements were conducted using 0.1 wt% for Dex500k and PEG6k, and 0.5 wt% for PEG1k, solutions supplemented with either NaCl or $MgCl_2$ (131-00161; FUJIFILM Wako Pure Chemical Co.) at a final concentration of 200 mM (see S5 for the details).

To quantify the degree of cation accumulation, we used a $Na^+$ probe, CoroNa Green Sodium Probe (C36675; Invitrogen), dissolved in dimethyl sulfoxide (DMSO; D0798; Tokyo Chemical Industry Co., Ltd., Tokyo, Japan). For comparison, NaCl was added to the PEG6k/Dex500k mixtures, both with and without λ DNA, to reach final concentrations ranging from 0 to 1 M. After isolating the coexisting PEG-rich and Dex-rich liquids, the $Na^+$ probe was then added (Fig.

3b). The fluorescence intensities were measured by using a confocal laser scanning microscope with excitation at 473 nm and emission detection of 490–590 nm (see S7 for the details).

**Conflicts of interest**

There are no conflicts to declare.

**Data Availability Statement**

The data that support the findings of this study are available from the corresponding author upon reasonable request.

**Table of Contents Graphics**

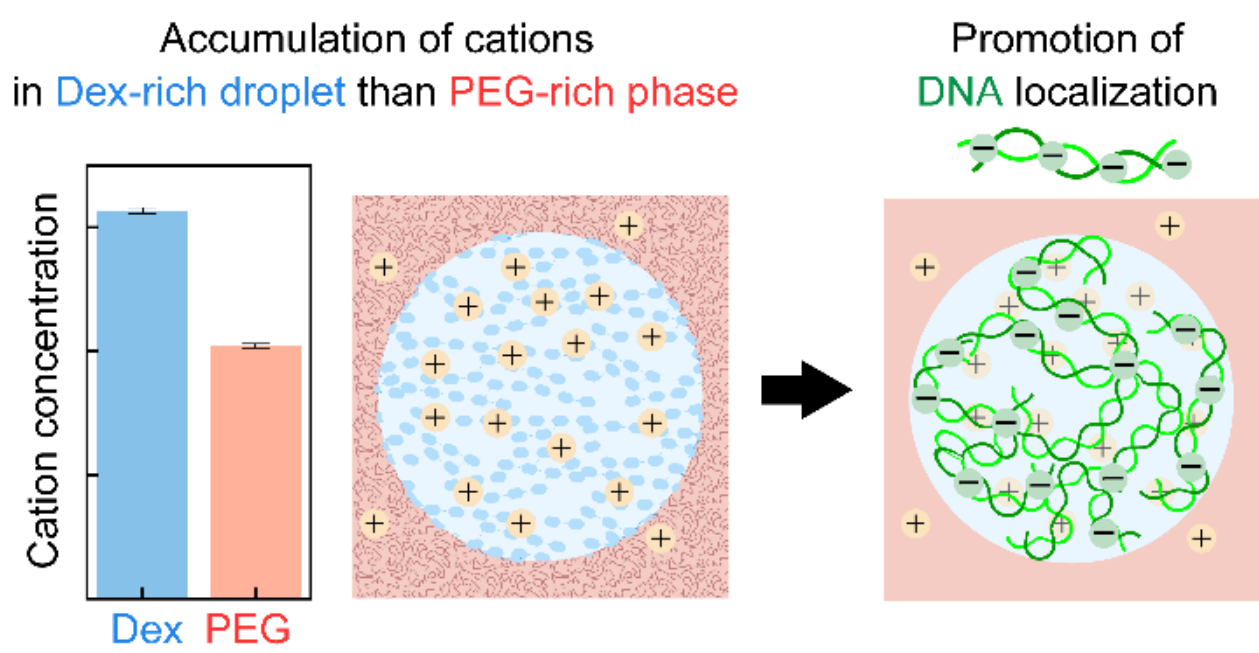

**Acknowledgements**

This research was partially funded by the Japan Society for the Promotion of Science (JSPS) KAKENHI (grant numbers 25K17355 (H. S.) 22H01188, 24H02287 (M. Y.)), JSPS Invitational Fellowship for Research in Japan (A. N.) and the Japan Science and Technology Agency (JST) (grant numbers FOREST, JPMJFR213Y; CREST, JPMJCR22E1(M. Y.)). A. N. also received support by the Deutsche Forschungsgemeinschaft (DFG, German Research Foundation) through Project no. 470113688.

# Supporting Information for

# Cation accumulation drives the preferential partitioning of DNA in an aqueous two-phase system

Hiroki Sakuta[1,2†], Yuki Akamine[1†], Akari Kamo[3], Hao Gong[1], Norikazu Ichihashi[1,2], Arash Nikoubashman[3,4,5*], Miho Yanagisawa[1,2,6*]

1 Komaba Institute for Science, Graduate School of Arts and Sciences, The University of Tokyo, Komaba 3-8-1, Meguro, Tokyo 153-8902, Japan
2 Center for Complex Systems Biology, Universal Biology Institute, The University of Tokyo, Komaba 3-8-1, Meguro, Tokyo 153-8902, Japan
3 Department of Physics, Graduate School of Science, The University of Tokyo, Hongo, 7-3-1, Bunkyo, Tokyo 113-0033, Japan
4 Leibniz-Institut für Polymerforschung Dresden e.V., Hohe Straße 6, 01069 Dresden, Germany
5 Institut für Theoretische Physik, Technische Universität Dresden, 01069 Dresden, Germany
6 Department of Mechanical Engineering, Keio University, Hiyoshi 3-14-1, Kohoku, Yokohama 223-8522, Japan

**AUTHOR INFORMATION**
†H.S. and Y.A. contributed equally to this paper.

**Corresponding authors**
E-mail: anikouba@ipfdd.de; myanagisawa@g.ecc.u-tokyo.ac.jp

**Table of contens**

## S1. Preparation of dsDNA

### S1. 1. DNA samples

Experiments were performed on double-stranded DNA (dsDNA) of lengths ranging from 50 bp to 49 kbp. λ DNA, as 49kbp DNA, was purchased from Nippon Gene Co., Ltd. (318-00414; Tokyo, Japan). As 50 bp, we used dsDNA fragments (NoLimits 50 bp DNA Fragment; SM1421; Thermo Scientific, Waltham, MA, USA).

### S1. 2. Restriction enzyme digestion

To obtain a 24 kbp fragment, λ DNA (New England Biolabs) was digested with XbaI (Takara Bio Inc.) at 37 °C for two hours. XbaI cleaves λ DNA at nucleotide position 24508, yielding fragments of 24508 bp and 23994 bp. After digestion, enzymes and buffer components were removed using column purification with the QIAquick PCR Purification Kit (QIAGEN N.V., Venlo, Netherlands). The lengths of dsDNA fragments were verified by 1% agarose gel electrophoresis. Agarose (Takara Bio Inc.) was dissolved in TAE buffer (20 mM acetic acid, 50 mM EDTA, 40 mM tris(hydroxymethyl)aminomethane, pH7.4. dsDNA samples were stained with SAFELOOK Load-Green (FUJIFILM Wako Pure Chemical Co.) and then subjected to electrophoresis in TAE buffer. The obtained gel images are shown in Fig. S2(a).

### S1. 3. DNA synthesis by polymerase chain reaction (PCR)

5 kbp, 2 kbp, 0.2 kbp, and 0.1 kbp dsDNA were synthesized by PCR using KOD One PCR Master Mix (KMM-101; Tokyobo Co., Ltd., Osaka, Japan), λ DNA (Klenow Fragment-treated; 318-05291; Nippon Gene Co., Ltd.) as a template, and primers. Primers comprised a standard forward primer (5′-GGTGCGAGTATCCGTACCATTC-3′) and sets of length-specific reverse primers designed to anneal at positions downstream on the λ DNA template as follows: 10kbp (3′-GCGCCGGTTCAGCAGAC-5′), 5 kbp (3′-CAGCCAGTCCGGCATCAATG-5′), 2 kbp (3′-TCGCTTACGTGGCATGCTG-5′), 200 bp (3′-GCTGGAGATCTGCCTCGC-5′), 100 bp (5′-CGGCGGCAGAGTCATAAAG-3′). PCR was performed in a thermal cycler using a reaction mixture containing template λ DNA (0.4 ng/μL), forward and reverse primers (0.5 pmol/μL each), and KOD One PCR Master Mix. PCR cycling was performed 30 cycles of 94 °C for 2 min, 98 °C for 10 sec, 68 °C for 3 min 20 sec on 10 and 5 kbp; 94 °C for 2 min, 98 °C for 10 sec, 68 °C for 3 min 20 sec on 2 kbp and 0.2 kbp and 0.1 kbp. After amplification by PCR, the dsDNA was purified using column filtration, and its length was confirmed by electrophoresis, as mentioned above. Gel images are shown in Fig. S2 (b) 10 kbp and 5 kbp; (c) 2 kbp; (d) 0.2 kbp and 0.1 kbp.

## S2. Preparation and observation of PEG/Dex mixtures with or without DNA

Stock solutions of PEG6k and Dex500k (20 wt%) were prepared by dissolving the polymers in distilled water at 60 °C. The solutions were then left to stand at room temperature (approximately 24 °C) for 1 day before mixing them to the final concentrations. The final concentrations of PEG, Dex, SYBR

Green I, Tris-HCl, 2-Mercaptoethanol, and dsDNA were 6 wt%, 2 wt%, 0.6X, 8 mM, 4%, and 10 ng/μL, respectively. The solutions were mixed using a vortex mixer for 7 seconds at 5000 rpm, with a minimum of two repetitions, to obtain Dex droplets with radii R ranging from 4 to 50 μm. The chamber consisted of two pieces of cover glasses (No. 1; 0.12–0.17 mm; Matsunami Glass Ind., Ltd., Osaka, Japan) attached with double-sided adhesive tape (J0410; Nitoms, Inc., Tokyo, Japan). Then, 15 μL of the solution was injected into a chamber and observed by a confocal microscope. Fluorescence images of the Dex droplets in a PEG solution were obtained using a confocal laser scanning microscope (IX83, FV1200; Olympus Inc., Tokyo, Japan), equipped with a water-immersion objective lens (UPLSAPO 60XW, Olympus Inc.). The PEG-rich phase with 0.2 wt% RB-PEG and dsDNA inside with SYBR Green I were excited at 473 and 559 nm wavelengths, respectively, and detected in the 490–590 and 575–675 nm ranges. The fluorescence intensity was analyzed using Fiji software (National Institute of Health (NIH), USA).

**S3. Estimation of polymer compositions in PEG/Dex mixtures**

To estimate the composition of each coexisting phase in the PEG/Dex mixture, with or without 100 mM NaCl, we measured the refractive index ($n$) and density ($\rho$) at ~25 °C. These measurements were taken using a reflectometer (Abbemat 3000, Anton Paar GmbH, Graz, Austria) with an accuracy of 0.0001 at 589 nm and a density meter (DMA 1001, Anton Paar GmbH) with a resolution of 0.05 mg/mL, respectively[1]. To reach thermal equilibrium of the PEG/Dex mixture, the PEG/Dex mixture was allowed to stand after centrifugation (10,000 rpm for 60 min at ~25 °C; Model 3780; Kubota Co., Tokyo, Japan) for 1 day. Afterward, we collected the upper PEG-rich and lower Dex-rich phases for the measurements. We initially measured the values of $\rho$ and $n$ for PEG/Dex mixtures with a 1:1 ratio, where the total concentration ranged from 0 to 150 mg/mL, both with and without 100 mM NaCl (Fig. S1). It shows that the presence of NaCl does not significantly alter the equilibrium compositions. Accordingly, we obtained these values for mixtures containing 2 wt% PEG and 6 wt% Dex. By comparing these values, we roughly estimated the composition of each coexisting phase; the PEG-rich and Dex-rich phases were determined to be 6 wt% PEG + 2 wt% Dex and 1 wt% PEG + 12 wt% Dex, respectively. Similarly, we obtained these values for mixtures containing 16 wt% PEG1k and 2 wt% Dex500k. By comparing these values with the mixtures of PEG6k and Dex500k at a 1:1 ratio, the Dex concentration in the Dex-rich phase will be approximately 20 wt%.

As for the Dex500k, the estimated radius of gyration $R_g$ and overlapping concentration $c^*$ are ~24 nm and ~5 wt%, respectively. By using these values, the mesh size ($\sim R_g(c^*/c)^{-3/4}$) is estimated to be ~12 nm for a 12 wt% Dex500k solution.[2-3]

**S4. Estimation of depletion interactions**

To assess the depletion contribution to the dsDNA accumulation, we estimate depletion interactions using the classical Asakura–Oosawa (AO) model. The radius of gyration for PEG6k, Dex500k, and dsDNA are taken as $R_{g,PEG}$ = 3 nm, $R_{g,Dex}$ = 24 nm, and $R_{g,DNA}$ = 500 nm, respectively[2, 4] (see also Fig.

S5). Approximate compositions is 12 wt% Dex500k + 1 wt% PEG6k in the Dex-rich phase and 6 wt% PEG6k + 2 wt% Dex500k in the PEG-rich phase. The depletion potential is defined as

$$U_{\mathrm{dep}} = -P_{\mathrm{osm}} \cdot V_{\mathrm{overlap}}$$

where the osmotic pressure $P_{osm} = \rho_i \cdot k_B T$ (i = PEG, Dex) is given by the polymer number density $\rho_i$ of PEG and Dex, Boltzmann's constant $k_B$, and absolute temperature $T$. The overlap volume $V_{overlap}$ corresponds to the excluded volume gained by crowding-induced aggregation. Since dsDNA is much larger than both PEG and Dex ($R_{g,DNA} >> R_{g,PEG}$ and $R_{g,Dex}$), we approximate $V_{overlap}$ using the flat plate contact model:

$$V_{overlap} = \pi R_{\mathrm{g,DNA}}^2 \cdot 2R_{\mathrm{g,i}} (\mathrm{i} = \mathrm{P}, \mathrm{D})$$

This yields $V_{overlap,PEG} \approx 4.7 \times 10^{-21}$ m$^3$ and $V_{overlap,Dex} \approx 3.8 \times 10^{-20}$ m$^3$, respectively.

For the PEG-rich aqueous phase ($\rho_{PEG} = 6\times10^{21}$ /m$^3$, $\rho_{Dex} = 0.24\times10^{20}$ /m$^3$), the depletion potential $U_{dep,\ aq} \sim -29\ k_B T$. For Dex-rich droplets ($\rho_{PEG} = 1\cdot10^{21}$ /m$^3$, $\rho_{Dex} = 1.4\cdot10^{20}$ /m$^3$), we obtain $U_{dep,drop} \sim -10\ k_B T$. The difference, $U_{dep,aq} - U_{dep,drop} \approx -19\ k_B T$, suggests that depletion would drive dsDNA into the Dex-rich droplets.

However, this AO-type model assumes that PEG and Dex act as mutually impenetrable hard spheres relative to dsDNA. For dsDNA, this assumption is particularly poor: The spherical volume defined by a λ 49 kbp DNA coil ($R_{g,DNA} = 500$ nm, $V = 5.2\times10^{-19}$ m$^3$) contains only a tiny fraction of actual DNA mass. Given a chain mass of $4.9\times10^{-17}$ g (assuming 620 g/mol per base pair) and DNA density of 1.7 g/cm³, the true occupied volume is only $2.9\times10^{-23}$ m$^3$, which is only ~0.006% of the coil volume. Thus, the model drastically overestimates excluded-volume interactions and neglects the highly penetrable, porous nature of dsDNA coils.

## S5. Estimate using Flory–Huggins theory

Within the framework of Flory–Huggins theory, we can formulate the free energy of a self-avoiding polymer chain in solution as:

$$\frac{F}{k_{\mathrm{B}}T} = \frac{R^2}{b^2 N} + \frac{1}{p} \cdot \frac{R^3}{b^3} \cdot (1 - \varphi) \cdot \log(1 - \varphi) \qquad (1)$$

where $R$ is the radius of the polymer, $b$ is the lattice constant, $N$ is the number of lattice sites occupied by one polymer (*i.e.*, dsDNA), $\varphi = N b^3/R^3$ is the fraction of lattice sites enclosed by the polymer, and $p$ is the number of lattice sites occupied by a solvent molecule (*i.e.*, water, PEG, or Dex). The first term in Eq. (1) represents the extension free energy, while the second one is the free energy contribution from solvent entropy. Here, we have assumed that the solvent molecules and polymers have no enthalpic interactions, so that $\chi = 0$. Further, Eq. (1) neglects the translational entropy of the polymer, since the chain's overall positional degrees of freedom contribute only an additive constant to the free energy and therefore do not affect the equilibrium coil size.

For $\varphi \ll 1$, we can expand $(1 - \varphi)\cdot\log(1 - \varphi) \approx -\varphi + \varphi^2/2$, so that the free energy becomes:

$$\frac{F}{k_{\mathrm{B}}T} \approx \frac{R^2}{b^2 N} + \frac{1}{p} \cdot \left( -N + \frac{1}{2} \frac{b^3 N^2}{R^3} \right) \qquad (2)$$

Notably, the term $F_{\mathrm{slv}} \equiv -k_{\mathrm{B}}TN/p$ corresponds to the extensive entropy (chemical-potential) contribution from removing $N/p$ solvent molecules to make room for the polymer and must be compensated to determine the physically relevant insertion free energy $\Delta F = F - F_{\mathrm{slv}}$. The last term in Eq. (2) is proportional to the usual mean-field excluded-volume energy $F_{\mathrm{excl}} = k_{\mathrm{B}}TN\varphi$.

The free energy $F$ as well as $\Delta F$ are minimized for:

$$R^* = \left(\frac{3}{4p}\right)^{1/5} bN^{3/5} \tag{3}$$

which has the same scaling as the standard expression for a self-avoiding chain in solution, except with an additional prefactor $\propto p^{-1/5}$. Thus, the polymers become less swollen with increasing solvent length $p$.

The corresponding free energy difference $\Delta F^* \equiv \Delta F(R^*)$ is then:

$$\frac{\Delta F^*}{k_{\mathrm{B}}T} \approx \left[\left(\frac{3}{4}\right)^{2/5} + \frac{1}{2}\left(\frac{4}{3}\right)^{3/5}\right]\frac{N^{1/5}}{p^{2/5}} \tag{4}$$

Thus, the free energy is lower in a molecular solvent ($N \gg p > 1$) than in an atomic solvent ($p = 1$) of the same volume, since replacing larger solvent molecules removes fewer translational degrees of freedom, thereby reducing the entropic penalty of polymer insertion.

## S6. Zeta potential measurement

Zeta potential measurements were performed using a zeta potential meter with a flat electrode cell chamber (ELSZneo; Otsuka Electronics Co., Ltd., Osaka, Japan). Dex15k (Mw = 15–25 kg/mol; *Leuconostoc* spp.; 31387; Sigma-Aldrich) was measured together with Dex500k, PEG6k, and PEG1k. For Dex500k and PEG6k, measurements were conducted using 0.1 wt% solutions supplemented with either NaCl or $MgCl_2$ at a final concentration of 200 mM. For Dex15k and PEG1k, 0.1 wt% solutions did not yield sufficient scattering intensity; therefore, 0.5 wt% solutions were prepared and measured under the same conditions. In contrast, for Dex500k and PEG6k, 0.5 wt% solutions generated excessively high scattering intensity. Representative zeta potential results are shown in Fig. 3(a) and Fig. S7. The particle size distributions obtained from dynamic light scattering (DLS) measurements performed simultaneously with zeta potential measurements are shown in Fig. S5. DLS measurements were carried out using the same zeta potential analyzer and chamber, employing backscattering detection at a scattering angle of 169°.

## S7. Quantification of $Na^+$ accumulation

We utilized an $Na^+$ probe, CoroNa Green Sodium Probe (C36675; Invitrogen), to examine the partition of $Na^+$ between the PEG-rich and Dex-rich phases. The probe was dissolved in DMSO at a concentration of 1 mM and stored at −20 °C. Prior to mixing it with the PEG/Dex mixtures, the probe was diluted to 100 μM with ultrapure distilled water and then added to achieve a final concentration of 10 μM. In the control condition, each coexisting phase of the PEG/Dex mixtures was isolated following centrifugation at 5000 rpm for 10 minutes at ~25 °C. NaCl was then added to each phase to reach a desired concentration with the $Na^+$ probe (Figs. S6, S8). For comparison, NaCl was added to the PEG/Dex mixtures, both with and without λ DNA, to reach final concentrations ranging from 0 to 1 M. After isolating the coexisting PEG-rich and Dex-rich liquids, the $Na^+$ probe was then added (Fig. 3b). The fluorescence intensities were measured by using a confocal laser scanning microscope with excitation at 473 nm and emission detection of 490–590 nm. It is important to note that when the $Na^+$ probe was added to the PEG/Dex mixture before isolation, it preferentially partitioned into the PEG-rich phase due to their higher affinity than Dex. To eliminate the interaction between the polymers and the $Na^+$ probe, the $Na^+$ probe was added to each phase after isolation.

## S8. Quantification of $Cl^-$ accumulation

Relative $Cl^-$ accumulation level was quantified using *N*-(Ethoxycarbonylmethyl)-6-Methoxy-quinolinium Bromide (MQAE), a chloride ion fluorescent probe. Typically, a PEG/Dex mixture was prepared according to the method described in S2, then centrifuged at 5000 rpm for 10 minutes at room temperature, allowing the PEG-rich and Dex-rich phases to fully separate into the upper and lower layers. Subsequently, they were aspirated into two PCR tubes with 10 μL volumes using 10 μL pipette tips. $Cl^-$ accumulation levels of the two phases were quantified using a fluorospectrometer (NanoDrop 3300, Thermo Fisher Scientific) by mixing the sample (1 μL) and aqueous $Cl^-$ probe solution (1 μL, 10 mM).

**Supplementary Figures**

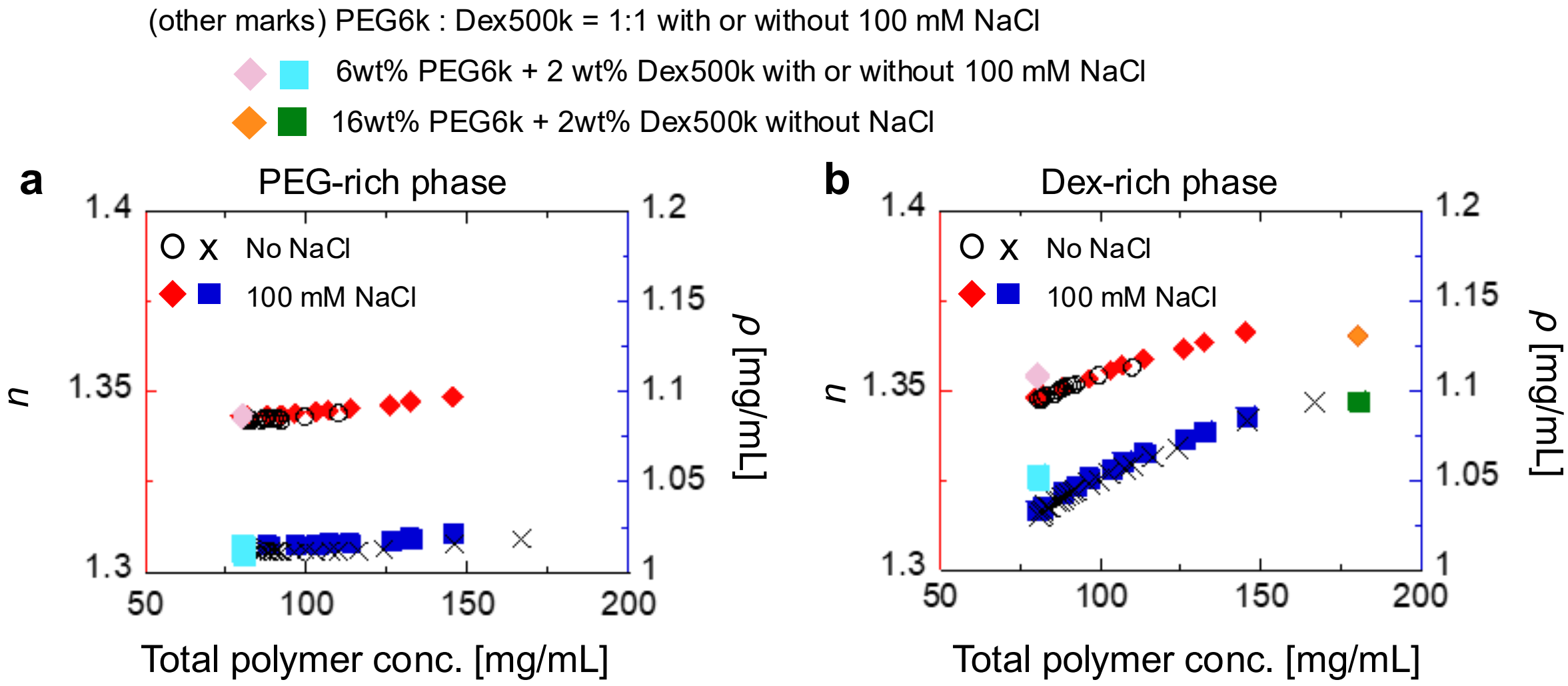

**Figure S1. Physical characteristics of PEG-rich and Dex-rich liquids.** The refractive index, $n$ (left $y$-axis), and density, $\rho$ (right $y$-axis), for (a) the PEG-rich phase and (b) the Dex-rich phase in mixtures of Dex500k and PEG6k (= 1:1) are shown against total polymer concentration. Symbols denote the presence (red and blue squares) or absence (open circles and crosses) of 100 mM NaCl. Pink diamonds and sky-blue squares indicate the specific experimental conditions shown in Fig. 2a (left) (*i.e.,* 6 wt% PEG6k and 2 wt% Dex500k solutions, with or without 100 mM NaCl). Orange diamonds and green squares indicate the specific experimental conditions shown in Fig. 2a (right) (*i.e.,* 16 wt% PEG1k and 2 wt% Dex500k solutions).

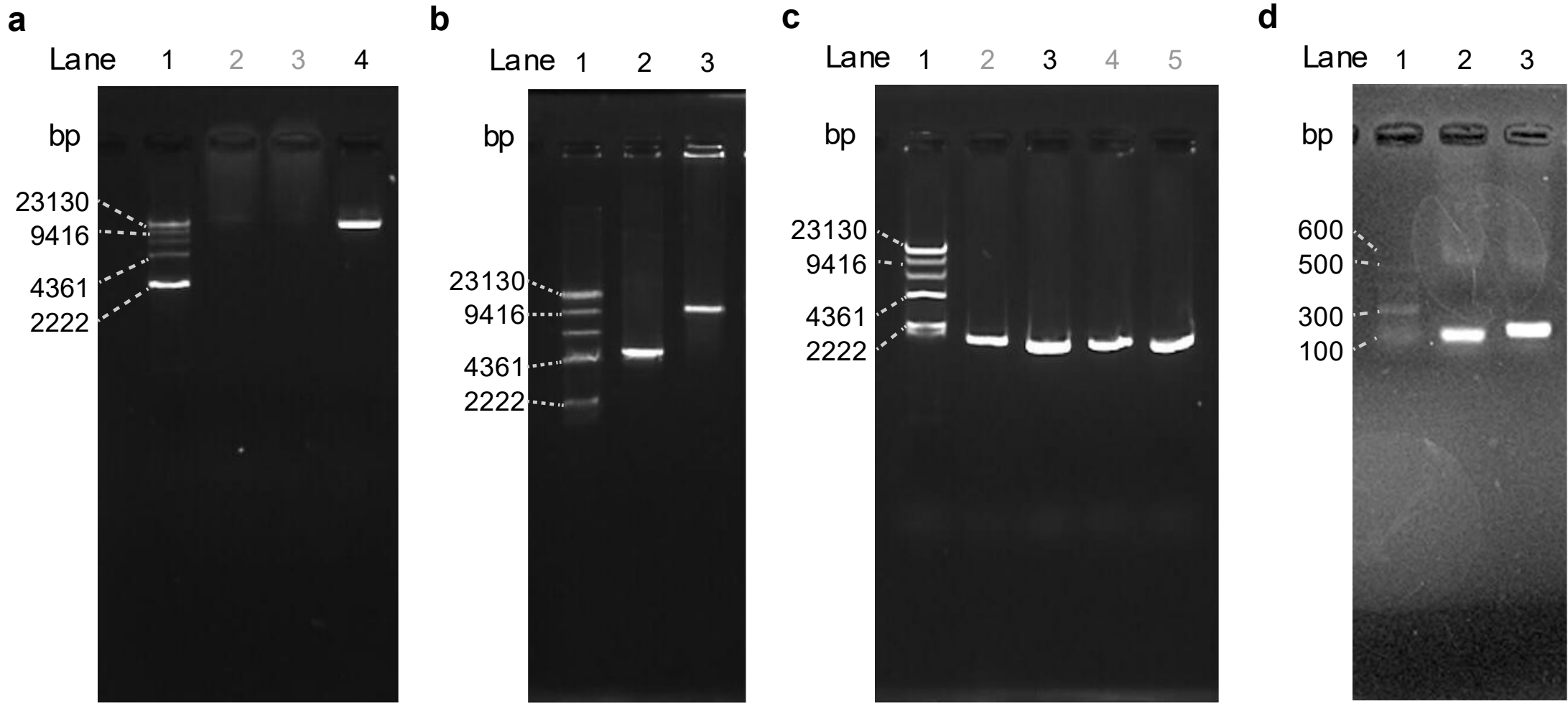

**Figure S2. Agarose gel electrophoresis of DNA fragments**. (a) 24 kbp fragments digested by the restriction enzyme (lane 4). Lane 1, ladder; lanes 2 and 3 are results of unsuccessful attempts at the synthesis of PCR with different primers. (b) DNA fragments of 10 kbp and 5 kbp amplified from λ DNA. Lane 1, DNA ladder; lane 2, 5 kbp; lane 3, 10 kbp. (c) 2 kbp fragments synthesized with various time sequences and primer sets. Lane 1, DNA ladder; lanes 2–5, 2 kbp fragments; lane 3, the 2 kbp DNA used for measurements. (d) Short DNA fragments. Lane 1, DNA ladder; lane 2, 0.1 kbp; lane 3, 0.2 kbp.

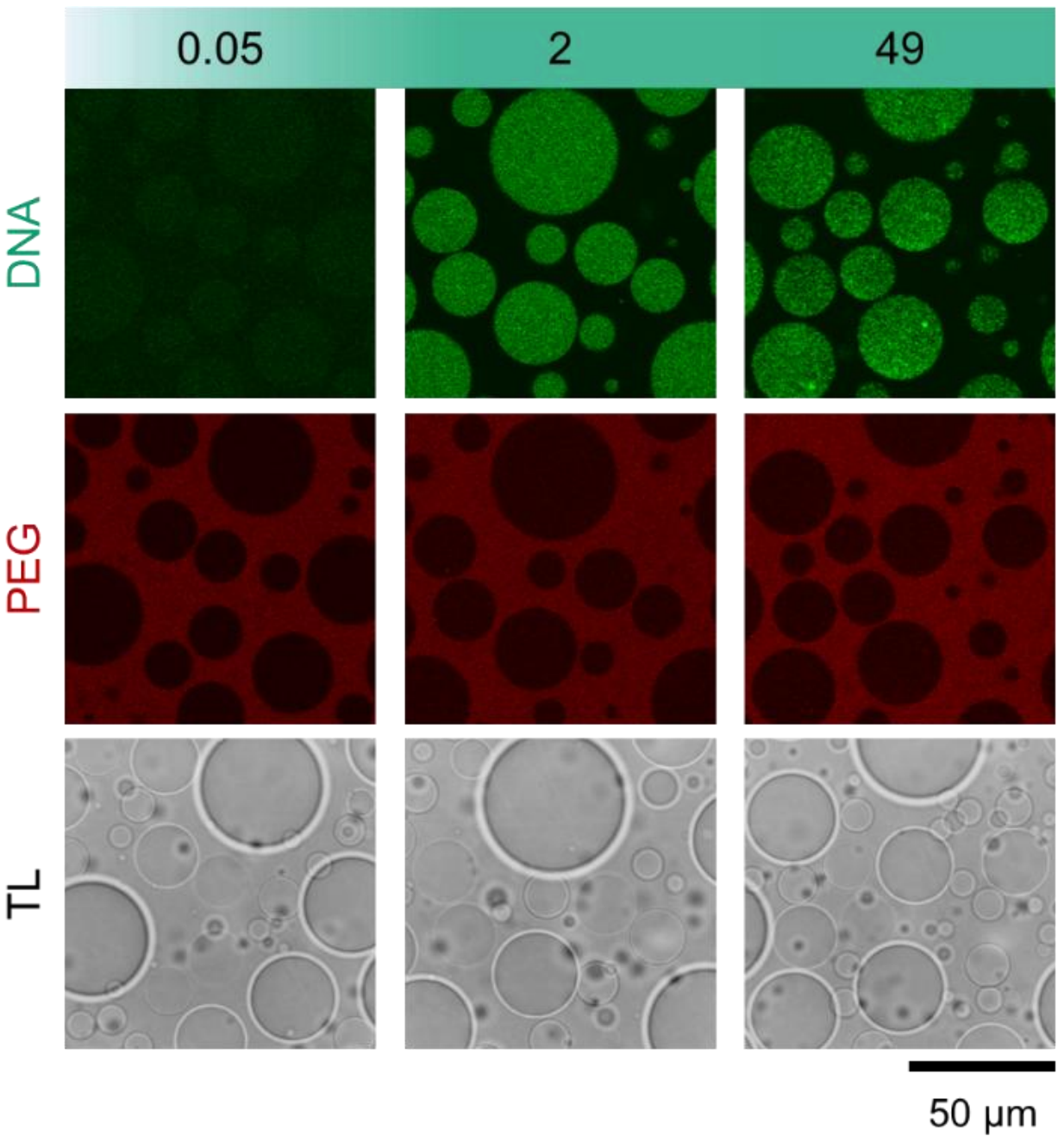

Figure S3. Transmitted light (TL) image and confocal fluorescence images of cross-sections of Dex-rich droplets containing dsDNA of different lengths. From left to right: 0.05 kbp, 2 kbp, and ~49 kbp λ DNA (green), imaged together with the PEG-rich phase labeled with RB–PEG (red).

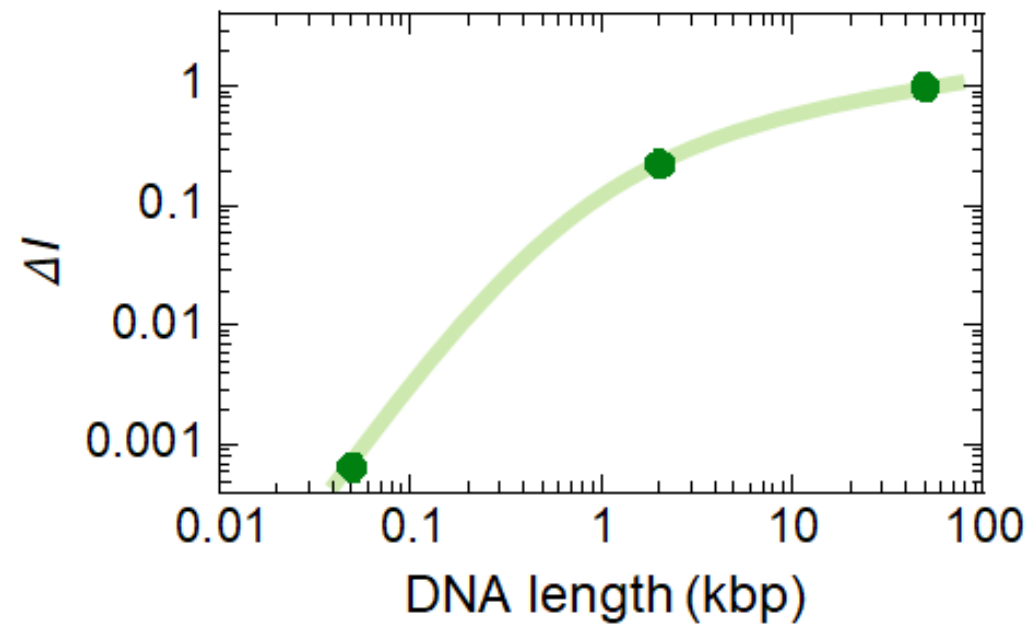

**Figure S4. Length-dependent accumulation of dsDNA within Dex-rich droplets in an aqueous PEG phase at a constant molar dsDNA concentration of 1 nM.** The normalized intensity difference between the Dex-rich and PEG-rich phases, $\Delta I$, is plotted as a function of dsDNA length. Error bars represent the standard error ($n \geq 51$ for each length). The solid green line is included as a visual guide. $\Delta I$ is approximately zero in the absence of dsDNA. A similar plot at a constant weight dsDNA concentration of 10 ng/μL is shown in Fig. 1g.

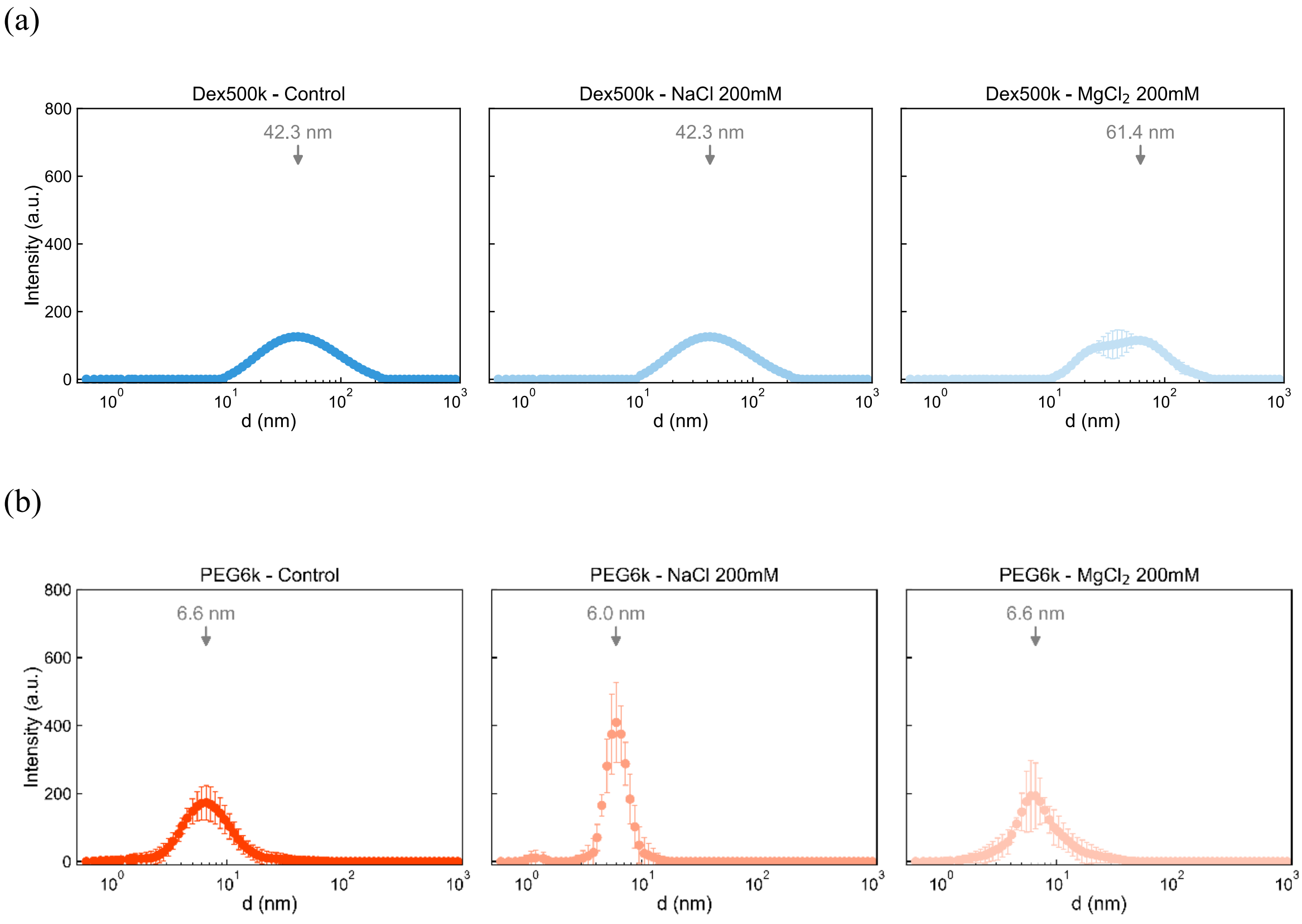

**Figure S5. Dynamic light scattering (DLS) analysis of polymer solutions.** (a) Hydrodynamic diameter, *d*, of Dex500k in water (Control), 200 mM NaCl, and 200 mM $MgCl_2$. The main peak positions are approximately at 42 nm (Control and NaCl) and 61 nm ($MgCl_2$). (b) The *d* of PEG6k in water (Control), 200 mM NaCl, and 200 mM $MgCl_2$. The main peak positions are 6-7 nm, regardless of the solution environment. Error bars indicate standard deviations ($N = 5$).

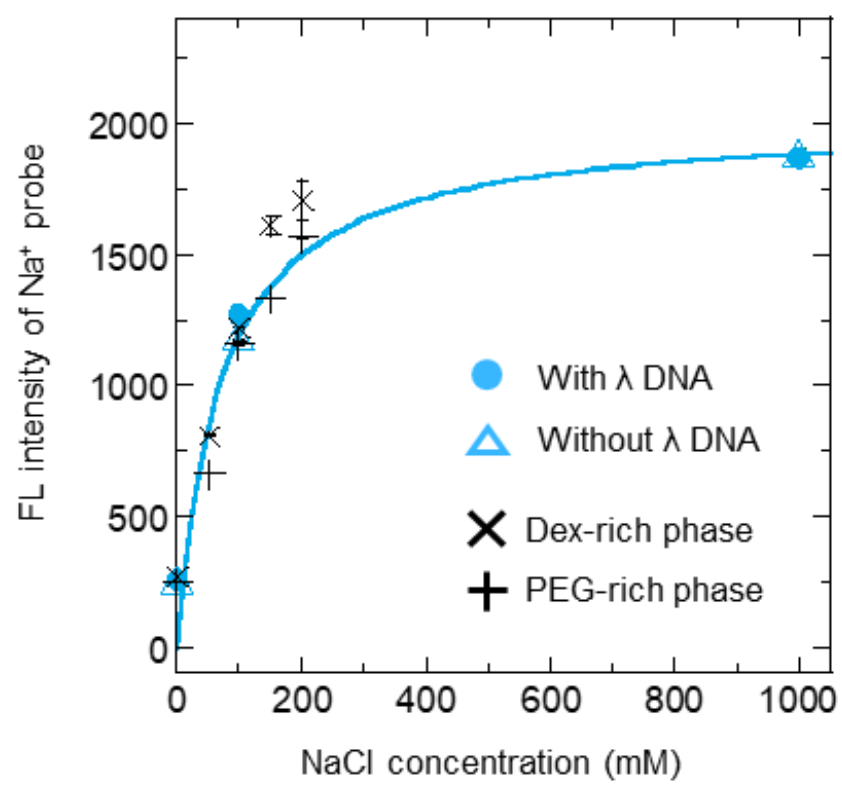

**Figure S6. Effect of dsDNA counterions on the $Na^+$ probe's intensity.** $Na^+$ probe's fluorescence intensities measured in NaCl solutions at various concentrations in the presence (blue crossed circles) or absence (blue triangles) of 10 ng/µL λ DNA. As controls, the intensities obtained from the isolated PEG-rich and Dex-rich liquids (NaCl and $Na^+$ probe were added after isolation; see experimental procedure in Fig. S8) are also shown (black crosses and plus signs). Error bars represent the standard deviation ($N \geq 3$).

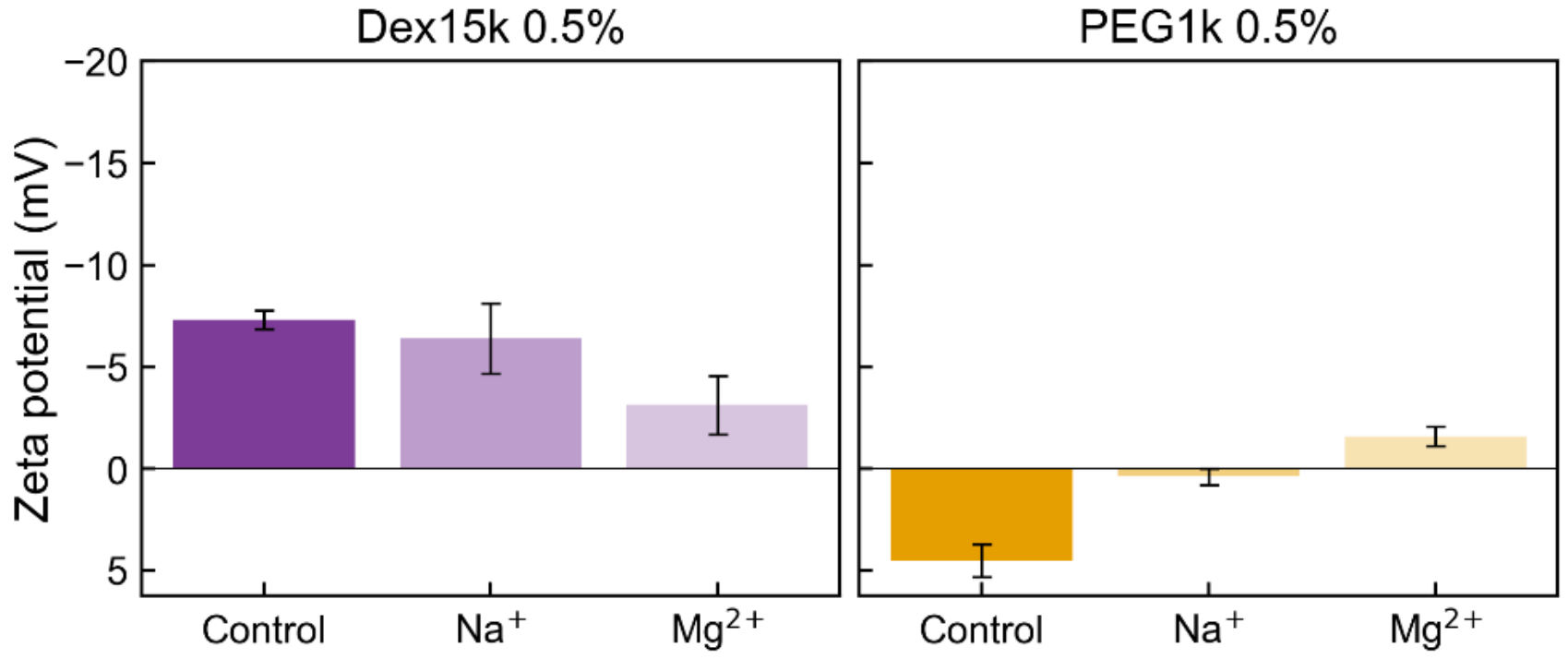

**Figure S7. Zeta potentials of short Dex15k and PEG1k.** Zeta potentials of (left) 0.5 wt% Dex15k and (right) 0.5 wt% PEG1k solutions are measured under different ionic conditions: in water (control, Ctrl), 200 mM NaCl ($Na^+$), and 200 mM $MgCl_2$ ($Mg^{2+}$). Error bars indicate standard deviations ($N = 3$).

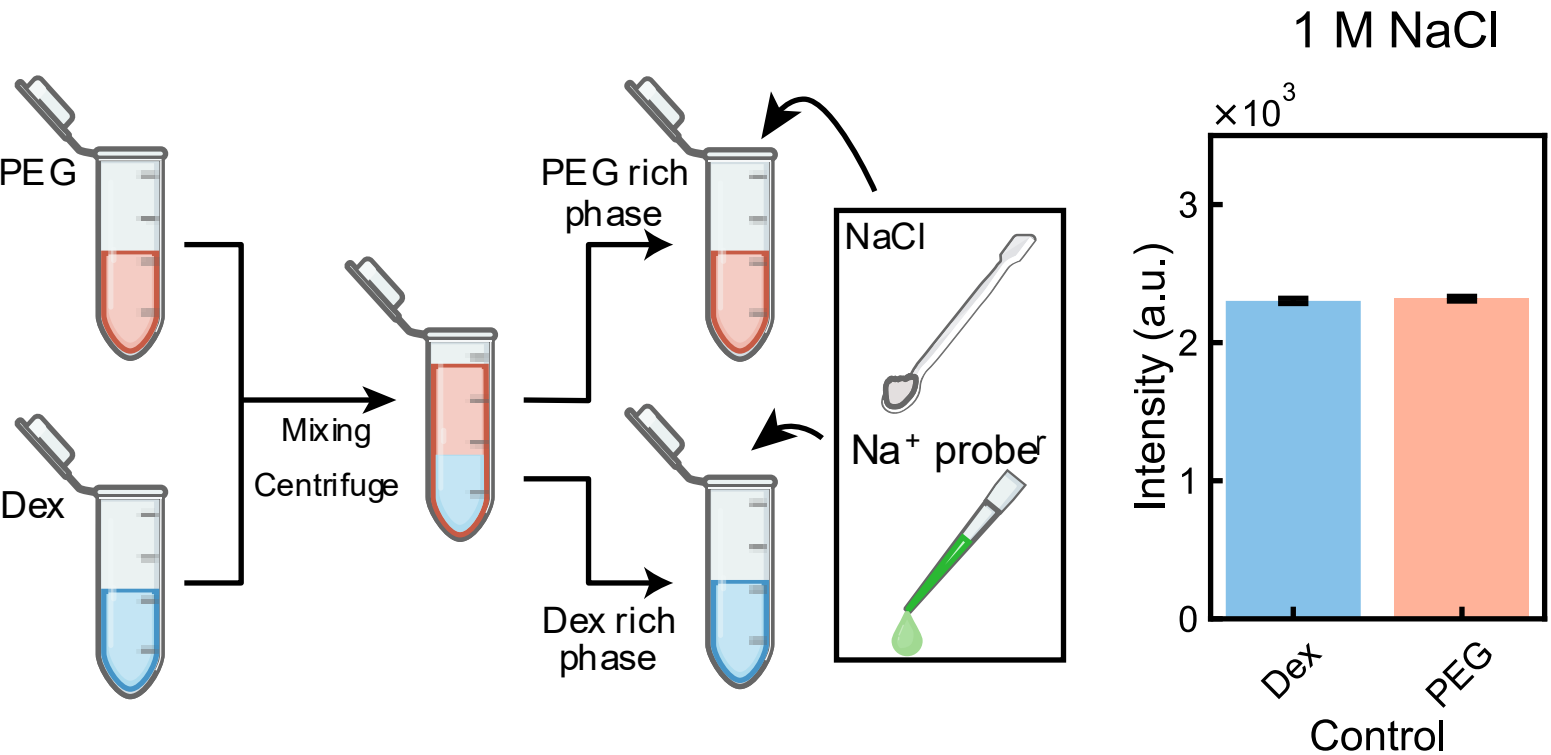

**Figure S8. Control experiment with salt added after isolation.** (left) After isolating the mixture into PEG-rich and Dex-rich liquids, $Na^+$ probe and NaCl were added. The final NaCl concentration is 1 M. (right) The $Na^+$ probe's intensity in the PEG-rich and Dex-rich phases, respectively. Because salt was added to both the PEG-rich and Dex-rich liquids after isolation, the $Na^+$ concentrations in each liquid, indicated by the $Na^+$ probes, are nearly identical, as expected. Error bars in the graph indicate standard deviation ($N = 3$).

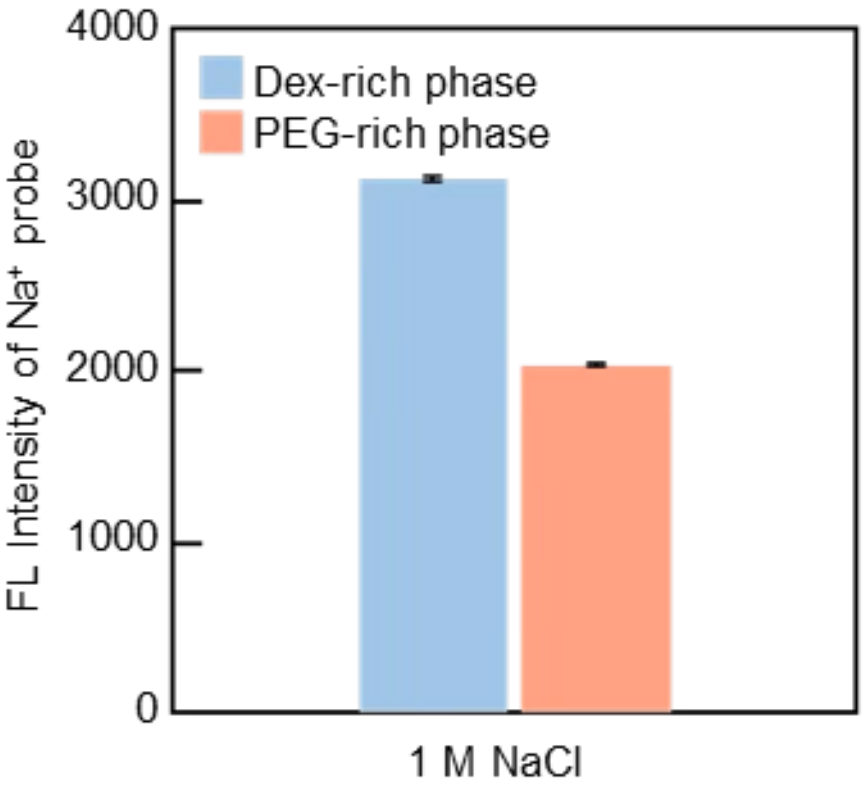

**Figure S9. The $Na^+$ probe intensity (a. u.) in each PEG- and Dex-rich phase in 1M NaCl added before isolation.** The error bars indicate standard deviation ($N = 3$).